\documentclass[aps,prl,superscriptaddress,amssymb,reprint]{revtex4-1}
\pdfoutput=1
\usepackage{amsmath, amsfonts}    
\usepackage{graphicx}   
\usepackage{verbatim}   
\usepackage{color}      
\usepackage{subfigure}  
\usepackage{hyperref}   
\usepackage{natbib}
\usepackage[USenglish]{babel}

\newcommand{\ket}[1]{\left|#1\right\rangle}

\newcommand{\up}{\uparrow}
\newcommand{\down}{\downarrow}
\newcommand{\uu}{\ket{\up\up}}

\newcommand{\dd}{\ket{\down\down}}

\newcommand{\mszero}{m_\mathrm s = 0}
\newcommand{\mspmone}{m_\mathrm s = \pm1}
\newcommand{\msmone}{m_\mathrm s = -1}

\begin{document}

\title{Heralded entanglement between solid-state qubits separated by 3 meters}
\author{H.~Bernien}
\author{B.~Hensen}
\author{W.~Pfaff}
\author{G.~Koolstra}
\author{M.S.~Blok}
\author{L.~Robledo}
\author{T.H.~Taminiau}
\affiliation{Kavli Institute of Nanoscience Delft, Delft University of Technology, P.O. Box 5046, 2600 GA Delft, The Netherlands}
\author{M.~Markham}
\author{D.J.~Twitchen}
\affiliation{Element Six Ltd., Kings Ride Park, Ascot, Berkshire SL5 8BP, United Kingdom}
\author{L.~Childress}
\affiliation{McGill University Department of Physics, 3600 Rue University, Montreal, QC H3A 2T8, Canada}
\author{R.~Hanson}
\email{r.hanson@tudelft.nl}
\affiliation{Kavli Institute of Nanoscience Delft, Delft University of Technology, P.O. Box 5046, 2600 GA Delft, The Netherlands}

\begin{abstract}
Quantum entanglement between spatially separated objects is one of the most intriguing phenomena in physics. The outcomes of independent measurements on entangled objects show correlations that cannot be explained by classical physics. Besides being of fundamental interest, entanglement is a unique resource for quantum information processing and communication. Entangled qubits can be used to establish private information or implement quantum logical gates~\cite{Nielsen2000,Raussendorf2001}. Such capabilities are particularly useful when the entangled qubits are spatially separated~\cite{Moehring2007,Ritter2012,Hofmann2012}, opening the opportunity to create highly connected quantum networks~\cite{Kimble2008} or extend quantum cryptography to long distances~\cite{Duan2001,Childress2006a}. Here we present a key experiment towards the realization of long-distance quantum networks with solid-state quantum registers. We have entangled two electron spin qubits in diamond that are separated by a three-meter distance. We establish this entanglement using a robust protocol based on local creation of spin-photon entanglement and a subsequent joint measurement of the photons. Detection of the photons heralds the projection of the spin qubits onto an entangled state. We verify the resulting non-local quantum correlations by performing single-shot readout~\cite{Robledo2011} on the qubits in different bases. The long-distance entanglement reported here can be combined with recently achieved initialization, readout and entanglement operations~\cite{Robledo2011,Neumann2010a,Neumann2008,Maurer2012,Pfaff2012} on local long-lived nuclear spin registers, enabling deterministic long-distance teleportation, quantum repeaters and extended quantum networks.
\end{abstract}

\maketitle

A quantum network can be constructed by using entanglement to connect local processing nodes, each containing a register of well-controlled and long-lived qubits~\cite{Kimble2008}. Solids are an attractive platform for such registers, as the use of nanofabrication and material design may enable well-controlled and scalable qubit systems~\cite{Ladd2010}. The potential impact of quantum networks on science and technology has recently spurred research efforts towards generating entangled states of distant solid-state qubits~\cite{Togan2010,Gao2012,DeGreve2012,Bernien2012a,Sipahigil2012,Patel2010,Flagg2010}.

A prime candidate for a solid-state quantum register is the nitrogen-vacancy (NV) defect centre in diamond. The NV centre combines a long-lived electronic spin (S=1) with a robust optical interface, enabling measurement and high-fidelity control of the spin qubit~\cite{Togan2010,Fuchs2009,DeLange2010,VanderSar2012}. Furthermore, the NV electron spin can be used to access and manipulate nearby nuclear spins~\cite{Robledo2011,Neumann2010a,Neumann2008,Maurer2012,Pfaff2012}, thereby forming a multi-qubit register. To use such registers in a quantum network requires a mechanism to coherently connect remote NV centres.

Here we demonstrate the generation of entanglement between NV centre spin qubits in distant setups. We achieve this breakthrough by combining recently established spin initialization and single-shot readout techniques~\cite{Robledo2011} with efficient resonant optical detection and feedback-based control over the optical transitions, all in a single experiment and executed with high fidelity. These results put solid-state qubits on par with trapped atomic qubits~\cite{Moehring2007,Ritter2012,Hofmann2012} as highly promising candidates for implementing quantum networks.

Our experiment makes use of two NV spin qubits located in independent low-temperature setups separated by 3 meters (Fig.~1a). We encode the qubit basis states $\ket{\up}$ and $\ket{\down}$ in the NV spin sublevels $\mszero$ and $\msmone$, respectively. Each qubit can be independently read out by detecting spin-dependent fluorescence in the NV phonon side band (non-resonant detection)~\cite{Robledo2011}. The qubits are individually controlled with microwave pulses applied to on-chip striplines~\cite{DeLange2010}. Quantum states encoded in the qubits are extremely long-lived: using dynamical decoupling techniques~\cite{DeLange2010} we obtain a coherence time exceeding 10$\,$ms (Fig.~1b), the longest coherence time measured to date for a single electron spin in a solid.

\begin{figure*}
	\includegraphics{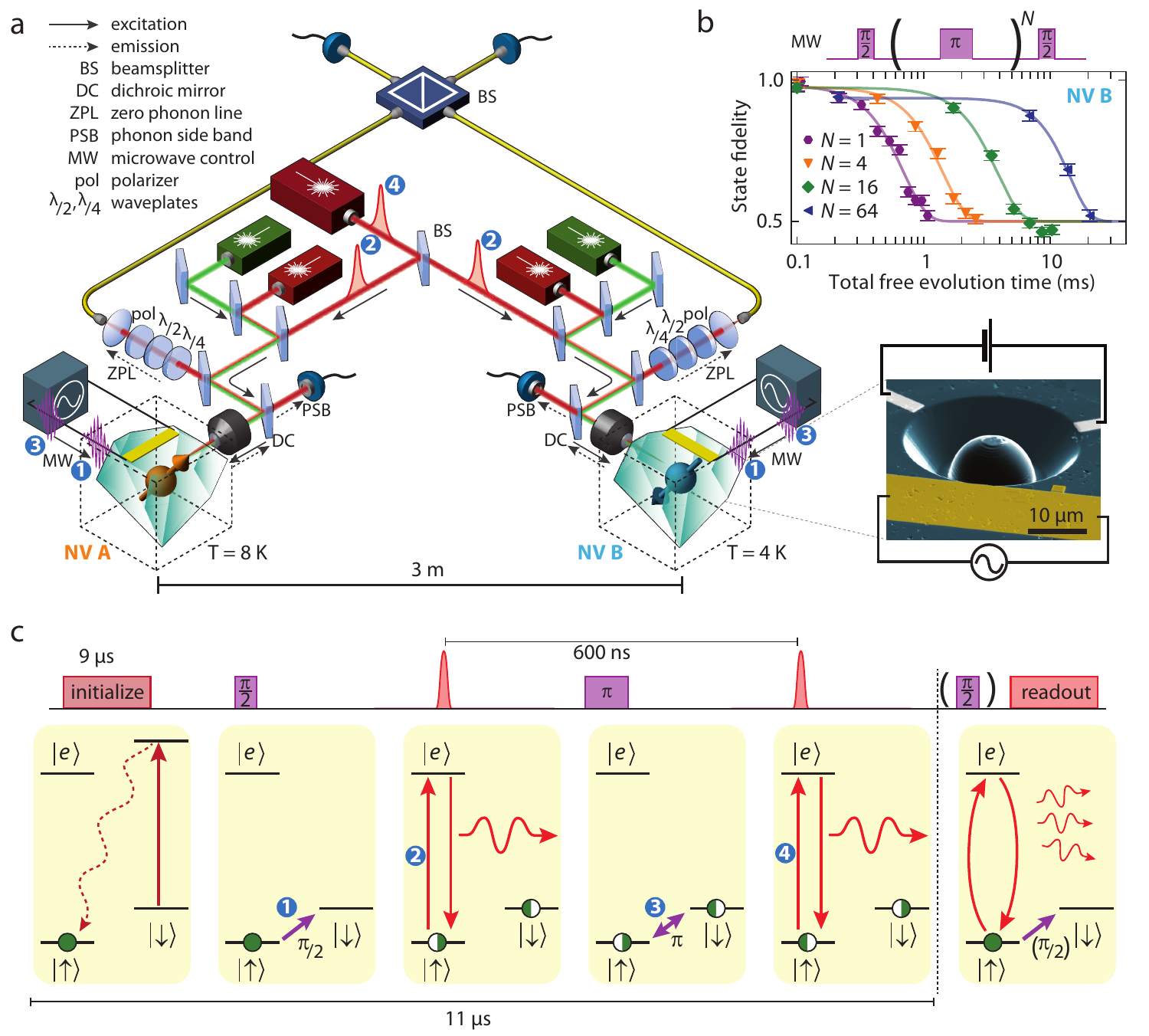}
	\caption{\label{LDE_fig1} Experimental setup and protocol for generating long-distance entanglement between two solid-state spin qubits. (a) Experimental setup. Each nitrogen vacancy (NV) centre resides in a synthetic ultrapure diamond oriented in the $\langle 111\rangle$ direction. The two diamonds are located in two independent low-temperature confocal microscope setups separated by 3 meters. The NV centres can be individually excited resonantly by a red laser and off-resonantly by a green laser. The emission (dashed arrows) is spectrally separated into an off-resonant part (phonon side band, PSB) and a resonant part (zero-phonon line, ZPL). The PSB emission is used for independent single-shot readout of the spin qubits~\cite{Robledo2011}. The ZPL photons from the two NV centres are overlapped on a fiber-coupled beamsplitter. Microwave pulses for spin control are applied via on-chip microwave striplines. An applied magnetic field of 17.5 G splits the $\mspmone$ levels in energy. The optical frequencies of NV~B are tuned by a d.c. electric field applied to the gate electrodes (inset, scanning electron microscope image of a similar device). To enhance the collection efficiency, solid immersion lenses have been milled around the two NV centres~\cite{Robledo2011}. (b) The coherence of the NV~B spin qubit as a function of total free evolution time $t_\text{FE}$ during an N-pulse dynamical decoupling sequence~\cite{DeLange2010}. Curves are fit to $A\exp[-(t_\text{FE}/T_\text{coh})^3]+0.5$. For N = 64 we find $T_\text{coh}=14.3\pm0.4\,$ms. (c) Entanglement protocol (details in main text), illustrating the pulse sequence applied simultaneously to both NV centres. Both NV centres are initially prepared in a superposition $1/\sqrt{2}(\ket{\up}+\ket{\down})$. A short $2\,$ns spin-selective resonant laser pulse creates spin-photon entanglement $1/\sqrt{2}(\ket{\uparrow 1}+\ket{\downarrow 0})$. The photons are overlapped on the beamsplitter and detected in the two output ports. Both spins are then flipped, and the NV centres are excited a second time. The detection of one photon in each excitation round heralds the entanglement and triggers individual spin readout.}
\end{figure*}

We generate and herald entanglement between these distant qubits by detecting the resonance fluorescence of the NV centres. The specific entanglement protocol we employ is based on the proposal of S. Barrett and P. Kok~\cite{Barrett2005}, and is schematically drawn in Figure~1c. Both centres NV~A and NV~B are initially prepared in a superposition $1/\sqrt{2}(\ket{\up}+\ket{\down})$. Next, each NV centre is excited by a short laser pulse that is resonant with the $\ket{\up}$ to $\ket{e}$ transition, where $\ket{e}$ is an optically excited state with the same spin projection as $\ket{\up}$. Spontaneous emission locally entangles the qubit and photon number, leaving each setup in the state $1/\sqrt{2}(\ket{\uparrow 1}+\ket{\downarrow 0})$, where 1 (0) denotes the presence (absence) of an emitted photon; the joint qubit-photon state of both setups is then described by $1/2(\ket{\uparrow_\text{A}\uparrow_\text{B}}\ket{1_\text{A}1_\text{B}}+\ket{\downarrow_\text{A}\downarrow_\text{B}}\ket{0_\text{A}0_\text{B}}+\ket{\uparrow_\text{A}\downarrow_\text{B}}\ket{1_\text{A}0_\text{B}}+\ket{\downarrow_\text{A}\uparrow_\text{B}}\ket{0_\text{A}1_\text{B}})$. The two photon modes, A and B, are directed to the input ports of a beamsplitter (see Fig.~1a), so that fluorescence observed in an output port could have originated from either NV centre. If the photons emitted by the two NV centres are indistinguishable, detection of precisely one photon on an output port would correspond to measuring the photon state $\ket{1_\text{A}0_\text{B}}\pm e^{-i\varphi}\ket{0_\text{A}1_\text{B}}$ (where $\varphi$ is a phase that depends on the optical path length). Such a detection event would thereby project the qubits onto the maximally entangled state $\ket{\psi}=1/\sqrt{2}(\ket{\uparrow_\text{A}\downarrow_\text{B}}\pm e^{-i\varphi}\ket{\downarrow_\text{A}\uparrow_\text{B}})$.

Any realistic experiment, however, suffers from photon loss and imperfect detector efficiency; detection of a single photon is thus also consistent with creation of the state $\uu$. To eliminate this possibility, both qubits are flipped and optically excited for a second time. Since $\uu$ is flipped to $\dd$, no photons are emitted in the second round for this state. In contrast, the states $\ket{\psi}$ will again yield a single photon. Detection of a photon in both rounds thus heralds the generation of an entangled state. The second round not only renders the protocol robust against photon loss, but it also changes $\varphi$ into a global phase, making the protocol insensitive to the optical path length difference~\cite{Barrett2005}. Furthermore, flipping the qubits provides a refocusing mechanism that counteracts spin dephasing during entanglement generation. The final state is one of two Bell states $\ket{\psi^\pm}=1/\sqrt{2}(\ket{\uparrow_\text{A}\downarrow_\text{B}}\pm\ket{\downarrow_\text{A}\uparrow_\text{B}})$, with the sign depending on whether the same detector ($+$), or different detectors ($-$) clicked in the two rounds.

A key challenge for generating remote entanglement with solid-state qubits is obtaining a large flux of indistinguishable photons, in part because local strain in the host lattice can induce large variations in photon frequency. The optical excitation spectra of the NV centres (Fig.~2a) display sharp spin-selective transitions. Here we use the $E_\text{y}$ transition (spin projection $\mszero$) in the entangling protocol and for qubit readout; we use the $A_1$ transition for fast optical pumping into $\ket{\up}$~\cite{Robledo2011}. Due to different strain in the two diamonds, the frequencies of the $E_\text{y}$ transitions differ by 3.5$\,$GHz, more than 100 linewidths. By applying a voltage to an on-chip electrode (Fig.~1a inset)  we tune the optical transition frequencies of one centre (NV~B) through the d.c. Stark effect~\cite{Bernien2012a,Bassett2011a} and bring the $E_\text{y}$ transitions of the two NV centres into resonance (Fig.~2a bottom).

\begin{figure*}
	\includegraphics{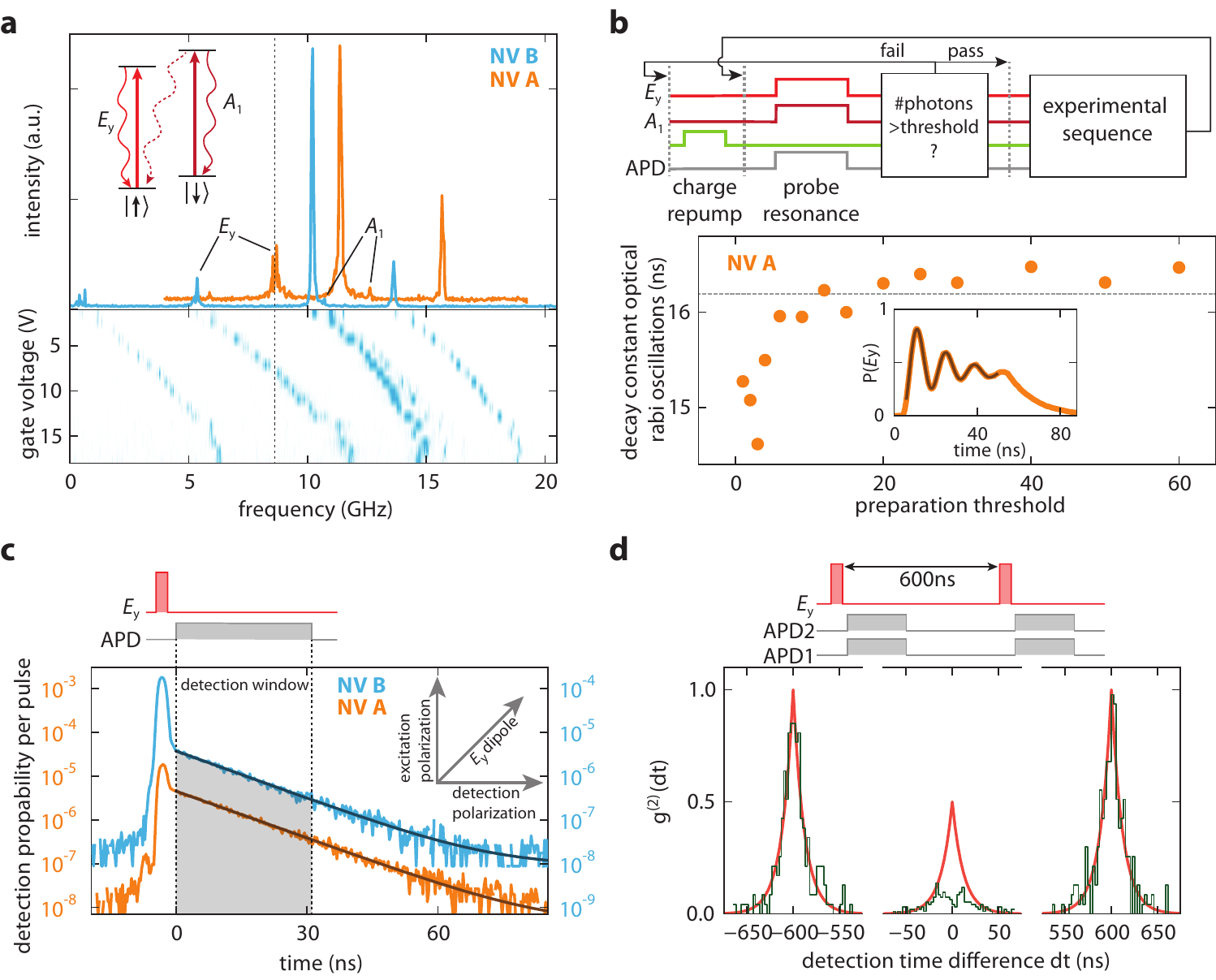}
	\caption{\label{LDE_fig2} Generating and detecting indistinguishable photons.
(a) Photoluminescence excitation spectra of NV~A and NV~B; frequency is given relative to 470.4515$\,$THz. Transitions are labeled according to the symmetry of their excited state. The $A_\text{1}$ transition is used to initialize the NV centre into the $\ket{\up}$ state ($\mszero$) and the $E_\text{y}$ transition is used for entanglement creation and single-shot readout. By applying a voltage to the gate electrodes of NV~B the $E_\text{y}$ transitions are tuned into resonance (dashed line). (b) Dynamical preparation of charge and optical resonance. Top: Preparation protocol. A 10$\,\mu$s green laser pulse (green line) pumps the NV centre into the desired negative charge state~\cite{Robledo2011}. Next the optical transition frequencies are probed by simultaneously exciting the $E_\text{y}$ and $A_\text{1}$ transitions for 60$\,\mu$s while counting the number of detected photons. Conditional on passing a certain threshold the experimental sequence is started (preparation successful) or else the protocol is repeated (preparation failed). APD, avalanche photodiode. Bottom: Line-narrowing effect of the preparation protocol exemplified by the dependence of the decay time of optical Rabi oscillations on preparation threshold. Dashed line indicates lifetime-limited damping~\cite{Robledo2010}. For the entanglement experiment we choose a threshold of 45 (20) photons for NV~A (NV~B). (c) Resonant optical excitation and detection. The polarization axis of the detection path is aligned perpendicular to the excitation axis. The dipole axis of the $E_\text{y}$ transition is oriented in between these two axes (inset). Remaining laser light reflection is time-filtered by defining a photon detection window that starts after the laser pulse. (d) Two-photon quantum interference using resonant excitation and detection. The $g^{(2)}$ correlation function is obtained from all coincidence detection events of APD~1 and APD~2 during the entanglement experiment. The sidepeaks are fit to an exponential decay; from the fit values, we obtain the expected central peak shape $g_\perp^{(2)}$ (red line) for non-interfering photons. The visibility of the interference is given by $(g_\perp^{(2)}-g^{(2)})/g_\perp^{(2)}$.}
\end{figure*}

Charge fluctuations near the NV centre also affect the optical frequencies. To counteract photo-ionization we need to regularly apply a green laser pulse to repump the NV centre into the desired charge state. This repump pulse changes the local electrostatic environment, leading to jumps of several linewidths in the optical transition frequencies~\cite{Robledo2010}. To overcome these effects, we only initiate an experiment if the number of photons collected during a two-laser probe stage (Fig.~2b) exceeds a threshold, thereby ensuring that the NV centre optical transitions are on resonance with the lasers. The preparation procedure markedly improves the observed optical coherence: as the probe threshold is increased, optical Rabi oscillations persist for longer times (see Fig.~2b). For high thresholds, the optical damping time saturates around the value expected for a lifetime-limited linewidth~\cite{Robledo2010}, indicating that the effect of spectral jumps induced by the repump laser is strongly mitigated.

Besides photon indistinguishability, successful execution of the protocol also requires that the detection probability of resonantly emitted photons exceed that of scattered laser photons and of detector dark counts. This is particularly demanding for NV centres since only about 3\% of their emission is in the zero-phonon line and useful for the protocol. To minimize detection of laser photons, we use both a cross-polarized excitation-detection scheme (Fig.~2c inset) and a detection time filter that exploits the difference between the length of the laser pulse (2$\,$ns) and the NV centreÕs excited state lifetime (12$\,$ns) (Fig.~2c). For a typical detection window used, this reduces the contribution of scattered laser photons to about 1\%. Combined with microfabricated solid-immersion lenses for enhanced collection efficiency (Fig.~1a inset) and spectral filtering for suppressing non-resonant NV emission, we obtain a detection probability of a resonant NV photon of about $4\times10^{-4}$ per pulse Ð about 70 times higher than the sum of background contributions.

The degree of photon indistinguishability and background suppression can be obtained directly from the second-order autocorrelation function $g^{(2)}$, which we extract from our entanglement experiment. For fully distinguishable photons, the value of $g^{(2)}$ would reach 0.5 at zero arrival time difference. A strong deviation from this behavior is observed (Fig.~2d) due to two-photon quantum interference~\cite{Hong1987} that, for perfectly indistinguishable photons, would make the central peak fully vanish. The remaining coincidences are likely caused by (temperature-dependent) phonon-induced transitions between optically excited states~\cite{Fu2009} in NV~A. The visibility of the two-photon interference observed here --- $(80\pm5)$\% for $|dt| < 2.56\,$ns --- is a significant improvement over previously measured values~\cite{Bernien2012a,Sipahigil2012} and key to the success of the entangling scheme.

To experimentally generate and detect remote entanglement, we run the following sequence: First, both NV centres are independently prepared into the correct charge state and brought into optical resonance according to the scheme in Figure~2b. Then we apply the entangling protocol shown in Figure~1c using a 600$\,$ns delay between the two optical excitation rounds. We repeat the protocol 300 times before we return to the resonance preparation step; this number is a compromise between maximizing the attempt rate and minimizing the probability of NV centre ionization. A fast logic circuit monitors the photon counts in real time and triggers single-shot qubit readout on each setup whenever entanglement is heralded, i.e. whenever a single photon is detected in each round of the protocol. The readout projects each qubit onto the \{$\ket{\up}$, $\ket{\down}$\} states (Z-basis), or on the \{$\ket{\up}+/-\ket{\down}$, $\ket{\up}-/+\ket{\down}$\} states (X or $-$X basis). The latter two are achieved by first rotating the qubit by $\pi/2$ using a microwave pulse before readout. By correlating the resulting single-qubit readout outcomes we can verify the generation of the desired entangled states. To obtain reliable estimates of the two-qubit state probabilities, we correct the raw data with a maximum-likelihood method for local readout infidelities. These readout errors are known accurately from regular calibrations performed during the experiment.

Figure~3 shows the obtained correlations. When both qubits are measured along Z (readout basis \{Z,Z\}), the states $\psi^+$ and $\psi^-$ (as identified by their different photon signatures) display strongly anti-correlated readout results (odd parity). The coherence of the joint qubit state is revealed by measurements performed in rotated bases (\{X,X\}, \{$-$X,X\}), which also exhibit significant correlations. Furthermore, these measurements allow us to distinguish between states $\psi^+$ and $\psi^-$. For $\psi^+$ the \{X,X\} (\{$-$X,X\}), outcomes exhibit even (odd) parity, whereas the $\psi^-$ state displays the opposite behavior, as expected. The observed parities demonstrate that the experiment yields the two desired entangled states.

\begin{figure*}
	\includegraphics{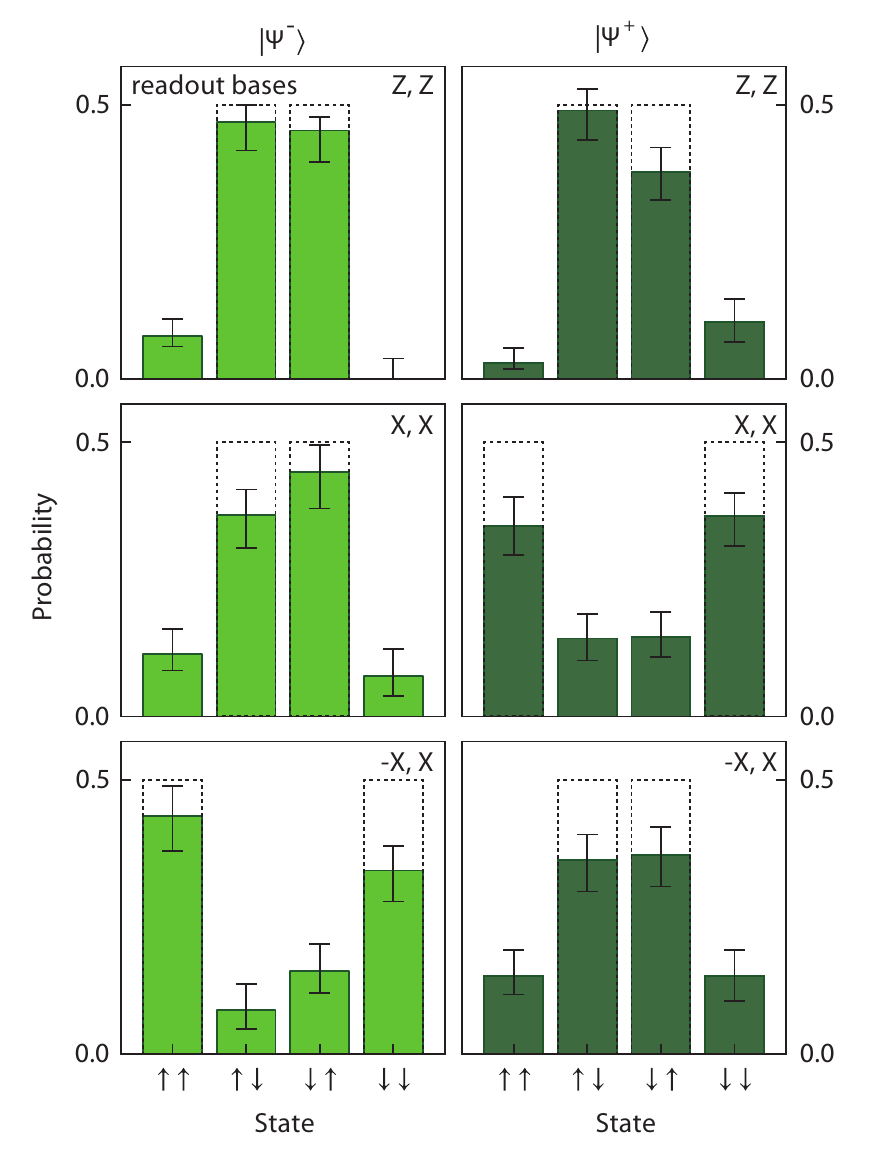}
	\caption{\label{LDE_fig3} Verification of entanglement: spin-spin correlations. Each time that entanglement is heralded the spin qubits are individually read out and their results correlated. The readout bases for NV~A and NV~B can be rotated by individual microwave control (see text). The state probabilities are obtained by a maximum-likelihood estimation on the raw readout results. Error bars depict 68\% confidence intervals; dashed lines indicate expected results for perfect state fidelity. Data is obtained from 739 heralding events. For $\psi^-$, the detection window in each round is set to 38.4$\,$ns, and the maximum absolute detection time difference $|\delta\tau|$ between the two photons relative to their laser pulses is restricted to 25.6$\,$ns. $\delta\tau=\tau_2-\tau_1$, where $\tau_1$ is the arrival time of the first photon relative to the first laser pulse and $\tau_2$ the arrival time of the second photon relative to the second laser pulse. For $\psi^+$ the second detection window is set to 19.2$\,$ns with $|\delta\tau|<12.8\,$ns, in order to reduce the effect of photo-detector afterpulsing.}
\end{figure*}

We calculate a strict lower bound on the state fidelity by combining the measurement results from different bases:
\begin{equation}
F = \langle\psi^\pm|\rho|\psi^\pm \rangle \geq \ 1/2(P_{\uparrow\downarrow}+P_{\downarrow\uparrow}+C)-\sqrt{P_{\uparrow\uparrow}P_{\downarrow\downarrow}},
\end{equation}
where $P_{ij}$ is the probability for the measurement outcome $ij$ in the \{Z,Z\} basis (i.e. the diagonal elements of the density matrix $\rho$) and $C$ is the contrast between odd and even outcomes in the rotated bases. We find a lower bound of $(69\pm5)$\% for $\psi^-$ and $(58\pm6)$\% for $\psi^+$, and probabilities of 99.98\% and 91.8\%, respectively, that the state fidelity is above the classical limit of 0.5. These values firmly establish that we have created remote entanglement, and are the main result of this paper.

The lower bound on the state fidelity given above takes into account the possible presence of coherence within the even-parity subspace \{$\uu$, $\dd$\}. However, the protocol selects out states with odd parity and therefore this coherence is expected to be absent. To compare the results to the expected value and to account for sources of error, we set the related (square-root) term in Eq. 1 to zero and obtain for the data in Figure~3 as best estimate $F=(73\pm4)$\% for $\psi^-$ and $F=(64\pm5)\%$ for $\psi^+$.

Several known error sources contribute to the observed fidelity. Most importantly, imperfect photon indistinguishability reduces the coherence of the state. In Figure~4a we plot the maximum state fidelity expected from photon interference data (Fig.~2d) together with the measured state fidelities, as a function of the maximum allowed difference in detection time of the two photons relative to their respective laser pulses. We find that the fidelity can be slightly increased by restricting the data to smaller time differences, albeit at the cost of a lower success rate (Fig.~4b).

\begin{figure*}
	\includegraphics{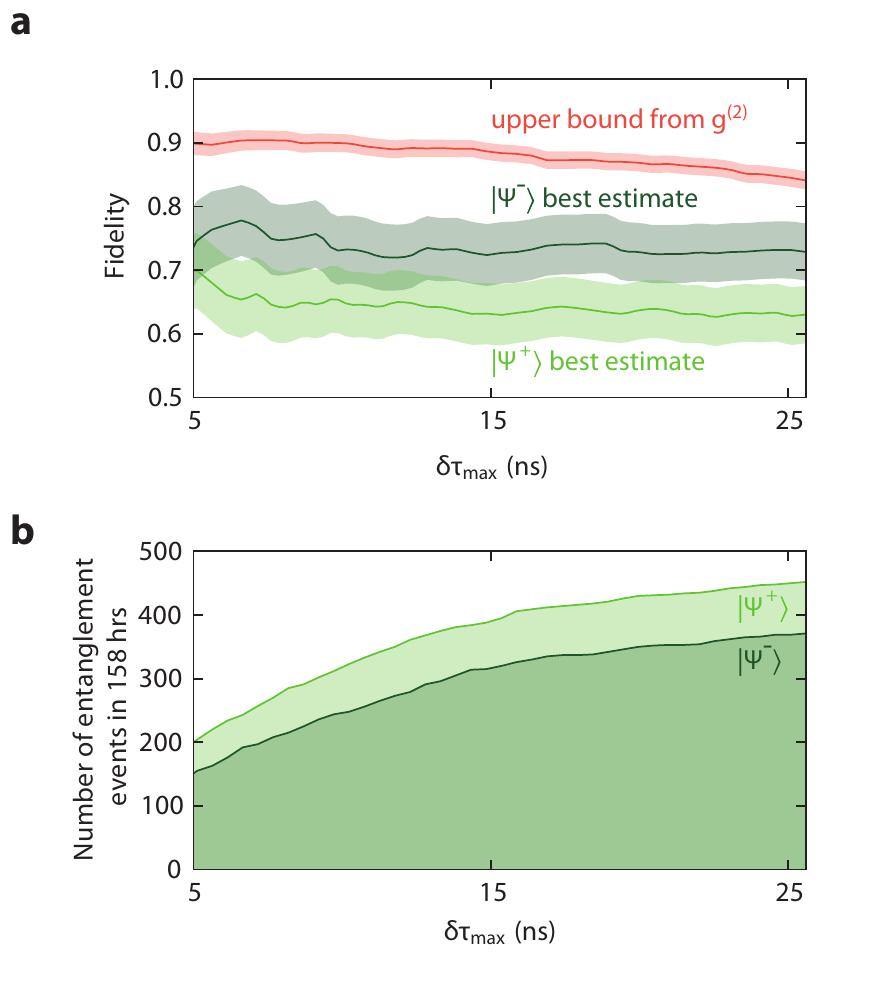}
	\caption{\label{LDE_fig4} Dependence of the fidelity and number of entanglement events on the detection time difference of the photons.
(a) Upper bound on the state fidelity from photon interference data and best estimate of the state fidelity from the correlation data as a function of the maximum allowed photon detection time difference ($|\delta\tau|<\delta\tau_\text{max}$). Detection time windows are chosen as in Figure~3. Shaded regions indicate 68\% confidence intervals. (b) Number of entanglement events obtained during 158 hours as a function of the maximum allowed photon detection time difference $\delta\tau_\text{max}$.}
\end{figure*}

The fidelity is further decreased by errors in the microwave pulses (estimated at 3.5\%), spin initialization (2\%), spin decoherence ($<1$\%) and spin flips during the optical excitation (1\%). Moreover, $\psi^+$ is affected by afterpulsing, whereby detection of a photon in the first round triggers a fake detector click in the second round. Such afterpulsing leads to a distortion of the correlations (see for example the increased probability for $\dd$ in Figure~3) and thereby a reduction in fidelity for $\psi^+$. Besides these errors that reduce the actual state fidelity, the measured value is also slightly lowered by a conservative estimation for readout infidelities and by errors in the final microwave $\pi/2$ pulse used for reading out in a rotated basis.

The fidelity of the remote entanglement can be significantly increased in future experiments by further improving photon indistinguishability. This may be achieved by more stringent frequency selection in the resonance initialization step and by working at lower temperatures, which will reduce phonon-mediated excited-state mixing~\cite{Fu2009}. Also, the microwave errors can be much reduced; for instance by using isotopically purified diamonds~\cite{Maurer2012} and polarizing the host nitrogen nuclear spin~\cite{Robledo2011}.

The success probability of the protocol is given by $P_\psi =1/2 \eta_\text{A}\eta_\text{B}$. $\eta_i$ is the overall detection efficiency of resonant photons from NV $i$ and the factor 1/2 takes into account cases where the two spins are projected into $\dd$ or $\uu$, which are filtered out by their different photon signature. In the current experiment we estimate $P_\psi \approx10^{-7}$ from the data in Figure~2c. The entanglement attempt rate is about 20$\,$kHz, yielding one entanglement event per 10 minutes. This is in good agreement with the 739 entanglement events obtained over a time of 158 hours. The use of optical cavities will greatly enhance both the collection efficiency and emission in the zero-phonon line~\cite{Aharonovich2011} and increase the success rate by several orders of magnitude.

Creation of entanglement between distant spin qubits in diamond, as reported here, opens the door to extending the remarkable properties of NV-based quantum registers towards applications in quantum information science. By transferring entanglement to nuclear spins near each NV centre, a nonlocal state might be preserved for seconds or longer~\cite{Maurer2012}, facilitating the construction of cluster states~\cite{Raussendorf2001} or quantum repeaters~\cite{Childress2006a}. At the same time, the auxiliary nuclear spin qubits also provide an excellent resource for processing and error correction. When combined with future advances in nanofabricated integrated optics and electronics, the use of electrons and photons as quantum links and nuclear spins for quantum processing and memory offers a compelling route towards realization of solid-state quantum networks.

\section{Acknowledgements}
We thank Fedor Jelezko, Pieter Kok, Mikhail Lukin, John Morton, Emre Togan and Lieven Vandersypen for helpful discussions and comments, and R.N.~Schouten and M.J.~Tiggelman for technical assistance. We acknowledge support from the Dutch Organization for Fundamental Research on Matter (FOM), the Netherlands Organization for Scientific Research (NWO), the DARPA QuASAR program, the EU SOLID and DIAMANT programs and the European Research Council through a Starting Grant.

\bibliographystyle{naturemag}
\bibliography{H_Bernien_ref.bib}

\end{document}